\documentclass[12pt]{article}
\usepackage{amssymb,amsfonts,amsmath,amsthm}
\usepackage[utf8]{inputenc}

\usepackage{graphicx}
\usepackage{amsmath, amssymb, amsfonts, amsgen,amstext,amsbsy, amsthm, epsfig, graphicx,rotating, subfigure}
\usepackage{amssymb,amsmath,epstopdf,graphicx}
\usepackage{appendix}
\usepackage{color, ulem, float}
\usepackage{url}
\usepackage{subfigure}
\usepackage{bbold}
\usepackage{soul}
\newtheorem{theorem}{Theorem}
\makeatletter \@addtoreset{theorem}{section}\makeatother

\newtheorem{lemma}[theorem]{Lemma}

\newtheorem*{theorem*}{Theorem}

\newtheorem{remark}[theorem]{Remark}

\newcommand{\figcaption}{\def\@captype{figure}\caption}
\newcommand{\tabcaption}{\def\@captype{table}\caption}
\makeatother

\newcommand{\beq}{\begin{eqnarray*}}
\newcommand{\eeq}{\end{eqnarray*}}
\newcommand{\beqn}{\begin{eqnarray}}
\newcommand{\eeqn}{\end{eqnarray}}

\def\be{\begin{equation}}
\def\ee{\end{equation}}
\def\bea{\begin{eqnarray}}
\def\eea{\end{eqnarray}}
\def\bd{\begin{displaymath}}
\def\ed{\end{displaymath}}
\def\bda{\begin{eqnarray*}}
\def\eda{\end{eqnarray*}}
\def\bsm{\begin{small}}
\def\esm{\end{small}}

\def\t0{\theta_0}

\def\ha1{\hat \beta_1}

\def\bnt{\begin{enumerate}}
\def\ent{\end{enumerate}}

\def\bsc{\begin{scriptsize}}
\def\esc{\end{scriptsize}}

\begin{document}

\title{Discovering the effect of nonlocal payoff calculation on the stabilty of ESS: Spatial patterns of Hawk-Dove game in metapopulations}

\author{Ozgur Aydogmus\thanks{Social Sciences University of Ankara, Department of Economics, H\"{u}k\"{u}met Meydani No:2 Ulus, Ankara, Turkey;
              E-mail: aydogmusozgur@gmail.com}
        }

\maketitle

\begin{abstract}
The classical idea of evolutionarily stable strategy (ESS) modeling animal behavior does not involve any spatial dependence. We considered a spatial Hawk-Dove game played by animals in a patchy environment with wrap around boundaries. We posit that each site contains the same number of individuals. An evolution equation for analyzing the stability of the ESS is found as the mean dynamics of the classical frequency dependent Moran process coupled via migration and nonlocal payoff calculation in 1D and 2D habitats. The linear stability analysis of the model is performed and conditions to observe spatial patterns are investigated. For the nearest neighbor interactions (including von Neumann and Moore neighborhoods in 2D) we concluded that it is possible to destabilize the ESS of the game and observe pattern formation when the dispersal rate is small enough. We numerically investigate the spatial patterns arising from the replicator equations coupled via nearest neighbor payoff calculation and dispersal.
\end{abstract}

\section{Introduction}
\label{intro}

Evolutionary game theory is a mathematically accessible way of describing different behavioral traits in a population each of which is associated with a pure strategy of the underlying game. The initial focus of evolutionary game theory was on the concept of evolutionarily stable strategies (ESS) which is used to enhance our understanding of the evolution of animal behavior by \cite{MSP,smith}. A strategy is defined to be an ESS if a small number of individuals playing a different strategy cannot invade a population playing it. An important question regarding ESS is if such a
strategy is attainable. The study by \cite{taylor0} extended the realm
of evolutionary game theory to include dynamics. In other words, they introduced the replicator equations relating the ESS concept with the equilibria of these equations \cite{HS}. Since then replicator equations are at the core of evolutionary game theory. This classical model describes the evolution of behavioral traits in an infinite population assuming that a given individual is equally likely to interact with any other. Key advances were made by relaxing some of the above mentioned assumptions. In particular,  the inclusion of finite populations and spatial structure in evolutionary games have accelerated the progress of the theory. 

Stochastic processes have been applied to model evolutionary game dynamics in finite populations \cite{finite}. There are a variety of microscopic rules describing the  game dynamics in finite populations such as birth-death update, death-birth update or pairwise comparison rules \cite{ohtsuki}. These update rules describes a class of Markov chains, transition probabilities of which are assumed to depend on frequencies of phenotypes and game parameters. For many of these processes, replicator equation is not only a limiting deterministic case \cite{traulsen} but also describes the mean dynamics of the underlying Markov chains.

Considering the fact that natural environments possess a spatial dimension,  meaning that individuals have limited mobility and interact with their neighbors, led many scholars to incorporate this important property into the study of evolutionary games. To analyse the effect of spatial structure on evolutionary game dynamics different approaches have been taken in to account: numerical simulations of games on grids (see {\it e.g.} \cite{NM,NBM,NS}) or more generaly on graphs \cite{allenowak,szabo}; and analytical  studies of replicator-diffusion equations (see, for example, \cite{ESS,Hof,Hof1,FM00}). We would like to note that a similar partial differential equation is found as a hydrodynamical limit of the frequency dependent Moran process \cite{macro}. In addition, integro-differential replicator equations taking nonlocal payoff calculation into consideration were obtained as a mesoscopic limit of the structured individual based models (see e.g. \cite{hwang,aydogmus}).

Here we let individuals play the Hawk-Dove game, originally
developed by \cite{MSP}, to describe certain scenarios in animal conflict modeling a contest over a shareable resource. There are two subtypes or morphs of one species with two strategies, Hawk and Dove. Both subtypes first display aggression. The Hawk escalates into a fight until it either wins or is injured (loses), whereas the Dove runs for safety if faced with major escalation. Unless faced with such escalation, the Dove attempts to share the resource. 

 Rather than taking the continuous space into account as in replicator-diffusion or integro differential equations, we consider a landscape ecology perspective and subdivide the environment into distinct but identical patches with periodic boundaries each of which contains a population of $K$ individividuals. We would like to note that stochastic population models considering a collection of patches on which a number of individuals  lives were studied by \cite{tilman,arrigoni}. In evolutionary game theory literature, a similar approach was taken by \cite{deme} to study a finite population of individuals subdivided into demes with the assumption of {\it local interactions} meaning that individuals interact only with other members of the same deme. 

In our dynamical setting we assume that a patch containing two subtypes of the same species is chosen randomly at each time step. Then an individual is drawn from this site with a probability depending on its fitness and replaces a randomly chosen individual from its natal site with probability $1-\mu$ or one of the neighboring sites. Such a process can be seen as a spatial coupling of frequency dependent Moran processes \cite{finite} via dispersal. Hence $\mu$ identifies the dispersal probability of a newborn. The fitness calculation, on the other hand, has two major components. The first one is the payoff calculation and the other is related to determining the effect of payoffs on fitness. 

Using replicator-diffusion equations as a modeling tool leads us to the assumption of local interactions ({\it i.e.} local payoff calculation) as in \cite{deme}. In the context of the Hawk-Dove game, this is to say that individuals living at a patch do not compete for the resources with the residents of other sites. This very same assumption, on the other hand, was criticized by \cite{main,doebelli} and relaxed by supposing that reproduction and hence population dynamics take place in a habitat patch whose resources are also used by individuals that live and reproduce in neighboring patches through foraging. This relaxation leads us to the fact that individuals from neighboring sites compete for the common resources and hence the payoff calculation for individuals residing in a patch does not depend solely on the local population configuration but also on the weighted average of the frequencies of the morphs in a certain neighborhood of the site. We would like to remark that there is a vast litrature on individual based models of evolutionary spatial games taking nonlocal payoff calculation in a neighborhood of a spatial location into account (see, for example, pioneering studies by \cite{NM,NBM}). The second major component of fitness calculation is related to the intensity of selection $w$, a parameter determining the strength of payoffs compared to the baseline fitness. As \cite{ohtsuki2} point out simple  as well as illuminating results arise in the limit of weak selection, $w\ll1.$ This is also the case for our model, hence we assume that the effect of payoffs is small when compared to baseline fitness.

The process considered here is a coupled system of Markov chains taking the spatial structure of the environment into account. This coupling between these chains is through dispersal and nonlocal payoff calculation. We find that the mean field dynamics of this coupled system of Markov chains is a coupled system of replicator equations (CRE). Here, we hypothesized that the magnitude of dispersal probability $\mu$ is comperable  to that of the small selection parameter $w\ll1.$ This hypothesis is shown to be a requirement to destabilize the ESS of the Hawk-Dove game. In particular, we analyse this limiting deterministic case near the ESS of the underlying Hawk-Dove game and find that small dispersal rate  gives rise to spatial pattern formation in 1D and 2D spatial regions with periodic boundaries when the magnitude of the dispersal probability is of order $O(w).$ We would like to note that spatial pattern formation is also possible in the strong selection regime. For a brief discussion and illustration of these patterns, see \ref{ap3}.

 The emergence of spatial inhomogeneity is related to the nonlocal (or quasi-local) interactions that are shown to be a mechanism for spatial pattern formation for a number of ecolgical processes \cite{main, pattern, aydogmus1,aydogmus3,killing2,doebelli,metanl,kisdi,utz}. In these earlier works it was shown that the spatial inhomegeniety is due to the fluctuating density of a species. Whereas, pattern formation in our model describes the fluctuations in the frequencies of the morphs due to the fact that replicator equations are used to model evolution in phenotype space.

The structure of this article is as follows: In Section \ref{mod}, we give a detailed description of our stochastic model and relate it to deterministic meanfield equations. In Section \ref{lsa}, we perform a linear stability analysis of the model and investigate the conditions for pattern formation. In Section \ref{ns}, we study the pattern formation numerically for the nearest neighbor interactions and investigate the effects of patch sizes and neighborhood types. Lastly, we discuss and summarize our findings in Section \ref{discussion}.

\section{From Coupled Moran Process to CRE}
\label{mod}

In this section we consider a class of two player symmetric games whose payof matrix is given as follows:

\[\begin{tabular}{ l | c r }
& A & B \\ \hline
A & $a$ & $ b$ \\ 
B & $c$ & $ d$ \\
\end{tabular}
\]

Here we consider a landscape ecology perpective and divide the habitat into identical distinct patches each of which contains $K$ individuals. Since our aim is to obtain and analyse the mean dynamics of a stochastic model, we assume that $K$ is large. We denote the set of these patches by $S.$ Each site $q$ in the set $S$ has a dispersal neighborhood denoted by $N^{q}.$ 

Suppose that each individual in the population is either type $A$ or $B.$ At each transition time, a site of origin $q\in S$ is chosen randomly, and the following actions take place in order: 
\begin{itemize}
\item An individual from the site $q$ is chosen to reproduce according to a frequency dependent probability
\item With probability $\mu,$ the offspring migrates to one of the neighbouring sites in $N^{q}$ equally likely and replaces a randomly chosen individual in this site.
\item The offspring replaces a randomly chosen individual from the site of origin $q$ with probability $1-\mu.$
\end{itemize}

Before proceeding to an introduction of our stochastic model, we discuss how to take a nonlocal payof calculation into account. Suppose that, for any site $q,$ the frequency of type $A$ individuals at time $t$ is given by $\mathcal X^{q}(t):=\mathcal X^{q}$ and denote the vector of these frequencies by $\mathbf{\mathcal X}=(\mathcal X^q)_{q\in S}.$ The payoff calculation is directly related to the foraging range of the species denoted by $\mathcal N^q.$ For the sake of simplicity we suppose that an individual from site $q$ is able to play the game with any individual in her site of origin $q$ or one of its neighboring sites in $ N^{q}$ and collects her payoff. Hence the foraging neighborhood may be taken as $\mathcal N^{q}= N^{q}\cup\{q\}.$ 
Then the frequency of type $A$ individuals in this neighborhood is given by \be\mathcal C^{q}(\mathbf{\mathcal X})=\frac{1}{|\mathcal N^{q}|}\sum_{r\in\mathcal N^{q}} \mathcal X^{r}\ee where $|\mathcal N^{q}|$ is used to denote the cardinality of set $\mathcal N^{q}.$ Hence payoffs of phenotypes $A $ and $B$ can be calculated as follows: 
\bea\label{payof}
\pi_A^{q}=a\,\mathcal C^{q}(\mathbf{\mathcal X})+b\big(1-\mathcal C^{q}(\mathbf{\mathcal X})\big) \text{ and }
\pi_B^{q}=c\,\mathcal C^{q}(\mathbf{\mathcal X})+d\big(1-\mathcal C^{q}(\mathbf{\mathcal X})\big).\eea We would like to remark that it is common to exclude self-interactions in calculating payoffs when finite population models of evolutionary game theory are taken into account. Nevertheless, our assumption that $K$ is large enables us to construct our model without this exclusion as done by many scholars (see, for instance, \cite{traulsen,macro,benaim}).

Using the selection parameter $w\in[0,1]$ to describe the degree to which payoffs contribute the fitnesses of these phenotypes leads us to the following fitness functions of type $A$ and $B$ individuals at site $q:$ 
\be f^{q}=1-w+w\pi_A^{q}\ee and \be g^{q}=1-w+w\pi_B^{q}\ee 
respectively. Hence, the frequency dependent probability to  choose an individual with strategy $A$ at site $q$ is given by 
$$\psi^{q}:=\psi^{q}(\mathcal X)=\frac{\mathcal X^{q}f^{q}}{\mathcal X^{q}f^{q}+(1-\mathcal X^{q})g^{q}}.$$ 
It is clear that the probability of  choosing a type $B$ individual is given by $\phi^{q}=1-\psi^{q}.$

Note that the selection parameter $w$  affects the contribution of payoffs to fitnesses of each type. In most studies a small selection parameter ($w\ll 1$) is used to simplify the analysis. This assumption is also biologically  reasonable  since there might be many other factors effecting the choice of the individual who is going to reproduce. Assuming small selection pressure leads us to the following approximation of the frequency dependent probabilty: $$\psi^{q}\approx\mathcal X^{q}\bigg[1+w(1-\mathcal X^{q})\big(\pi_A^{q}-\pi_B^{q}\big)\bigg].$$ 

Now we are ready to set up our Markov process. Suppose that transition times of coupled Moran process is given by $\mathbb T=\{0,\tau,2\tau,\cdots\}$ where $\tau=1/K.$  denotes its transition probabilities as follows:
$$P^{(q)}_{k,k\pm1}=P\Big(\mathcal X^q(t+\tau)=\mathcal X^q\pm\frac{1}{K}| \mathcal X^q(t)=\mathcal X^q\Big).$$ 
It is now possible to construct the transition probabilities of the Markov process. At site $q$ an increase in the number of type $A$ individuals is possible only if an individual of type $B$ is chosen from the site $q$ and replaced by a type $A$ individual.  The probability of chosing a type $A$ individual from the origin site is $\psi^q.$ This individual replaces a type $B$ individual with probability $(1-\mu)\psi^q\cdot(1-\mathcal X^q)$ in site $q.$ Or an individual of type $A$ can be chosen from the neighborhood $N^q$ to reproduce and disperse to site $q$ with probability $\sum_{r\in N^q}\frac{\mu}{s(N^r)} \mathcal \psi^r$ and replaces a type $B$ individual from site $q.$ In a similar way, one can construct the probability of an increase in the number of type $B$ individuals. Since we are dealing with a population of populations on a grid, we assume that each site has the same number of neighboring sites {\it i.e.} $s=|N^{q}|$ for any $q\in S.$ In the following section, we specify a few neighborhoods to study the stability of the mixed strategy ESS of the Hawk-Dove game.

Here we hypothesize that the magnitude of the dispersal probability $\mu$ is comperable with that of the selection parameter {\it i.e.} $\mu=\delta\cdot w$ for some positive constant $\delta.$ This weak dispersal hypothesis is required to simplify the equations governing the population dynamics and to destabilize the space homogenous ESS of the game in the weak selection regime. The former simplification is due to the fact that $\frac{\mu}{s}\sum_{r\in N^q} \mathcal \psi^r=w\frac{\delta}{s}\sum_{r\in N^q} \mathcal \mathcal X^r+O(w^2).$ For a justification of the latter, see Remark in Section \ref{remark}. Destabilization of the ESS of the Hawk-Dove game is also  possible (see \ref{ap3}) when the magnitude of dispersal probability $\mu$ is comperable to that of the selection parameter $w$ which is not assumed to be small. Consequently, we have the following transition probabilities in the weak selection regime:

$$P^{(q)}_{k,k+1}=\biggl(\psi^{q}+w\frac{\delta}{s}\Bigl( \sum_{r\in N^q}\mathcal X^r - s\mathcal X^q\Bigr)\biggr)(1-\mathcal X^q)+O(w^2)$$ 
$$P^{(q)}_{k,k-1}=\biggl(\phi^{q}-w\frac{\delta}{s}\Bigl( \sum_{r\in N^q}\mathcal X^r - s\mathcal X^q\Bigr)\biggr)\mathcal X^q+O(w^2)$$ and $$P^{(q)}_{k,k}=1-P^{(q)}_{k,k+1}-P^{(q)}_{k,k-1}.$$ 

Let $\mathbf x=(x_q)_{q\in S}$ be a solution to coupled replicator equations (CRE) that is given as follows: 
\be\label{general}\dot x_q= {\delta}w( \Delta x)_q+ wx_q(1-x_q) (\pi_A^{q}-\pi_B^{q})\ee where the discrete dispersal operator $\Delta$ is defined by the following equality:
\beq \label{1D} (\Delta \mathbf x)_q:=\frac{1}{s}\sum_{r\in N_q}x_r - x_q\eeq for any $q\in S.$ 

To connect the above given discrete time Markov chain with a continuous time differential equation we need to consider a piecewise affine interpolation of the process at any time $t\in[n\tau,(n+1)\tau]$ as follows: 
\beq\hat{\mathbf{\mathcal X}}(t)={\mathbf{\mathcal X}}(n\tau)+\frac{t-n\tau}{\tau}\big({\mathbf{\mathcal X}}(n\tau+\tau)-{\mathbf{\mathcal X}}(n\tau)\big)\eeq for any $n\in\mathbb N.$ Connection between the above defined interpolated Markov chain and CRE \eqref{general} is given by the following lemma. 

\begin{lemma}\label{lem} For any finite time $T>0$ there is a scalar $c=c(T)$ such that\beq P\bigg(\max_{0\leq t\leq T}\max_{q\in S} \big\{|\hat{\mathcal X}^q-x_q|:q\in S\big\}\geq\varepsilon:\mathbf{\mathcal X}(0)=\mathbf x(0)\bigg)\leq2|S|e^{-2\varepsilon c K}\eeq for any $\varepsilon>0$ and all $\mathbf x(0) \in [0,1]^{|S|}.$
\end{lemma}

The above given result can be proved following the proof of \cite[lemma 1]{benaim} which only requires replacing the vector field used in their paper with the Lipchitz continuous vector field associated with equation \eqref{general}. The result implies that the trajectories of the Markov chain and equation \eqref{general} deviate from each other with a probability that exponentially approaches to zero as $K$ increases. In particular, the above given bound can be used to obtain a strong law of large numbers result for any finite time $T$ as follows: \beq\lim_{K\to\infty} P\bigg(\max \big\{|{\mathcal X}^q(n\tau)-x_q(n\tau)|:n\tau<T,q\in S\big\}\geq\varepsilon:\mathbf{\mathcal X}(0)=\mathbf x(0)\bigg)=0 \text{ a.s.}\eeq 
which mimics the result obtained by \cite{kurtz} for continuous time density dependent Markov chains.

\section{Stability of Mixed Strategy ESS of the Hawk-Dove Game }
\label{lsa}

Recall the classic Hawk-Dove game involving two discrete strategies. Individuals playing a Hawk strategy always fight to injure or kill their opponent.
Whereas individuals employing the Dove strategy always display and never escalate the contest to serious fighting. If two individuals meet
and both adopt the Hawk strategy, at least one will be seriously injured in the contest. Similarly, if two players both adopt the Dove strategy, there is some
cost to continued displaying. When one player adopts a Hawk strategy and the other plays Dove, the Hawk wins the contested resource. 

The payoff matrix of a Hawk-Dove game satisfies $a<c$ and $d<b.$ To simplify the notation we introduce the following parameters 
\be \label{ab}\alpha=a-b-c+d \text{ and } \beta=d-b.\ee 
In terms of these two parameters above the given condition can be repeated as $\alpha,\beta<0$ and the mixed strategy ESS of the game is given by $e=\frac{\beta}{\alpha}.$ Hence, we denote the spatially homogenous solution of the CRE in vector form $\mathbf e=(e)_{q\in S}.$ 

We study the stabiliy of mixed strategy ESS of the Hawk-Dove game under CRE. In Section \ref{mod}, we did not specify the neighborhoods or the spatial dimension of the habitat. Here we first describe the simple nearest neighbor dispersal and interaction in 1D and 2D spatial habitats used in both ecological studies by \cite{doebelli, utz,kisdi,metanl,cosner12}, and game theoretical studies initiated by \cite{NBM,NM}. Then we give a general form of the mean field equation and find the conditions under which the ESS can be destabilized.

\subsection{Calculating Eigenvalues in 1D Habitat}\label{eig1}
Consider a one dimensional periodic habitat {\it i.e.} $S=\{0,1,2,\cdots,Q-1\}$ which forms a ring. Such a fragmented habitat may be reasonable for a species living around a lake or a pond. 
The nearest neighborhood interaction (or payoff calculation) makes sense if the foraging behavior of animals is taken into account. Thus, we consider the following dispersal and interaction neighborhoods:
\be N_1^{q}= \{q-1,q+1\} \text{ and } \mathcal N_1^q =\{q-1,q,q+1\}\ee 
The former neighborhood automatically identifies a one dimensional discrete dispersal operator:
\be(\Delta_1\mathbf x)_q=\frac{1}{2}(x_{q-1}+x_{q+1})-x_q\ee 
which is widely used as discretization of the diffusion term in reaction-diffusion equations (see {\it e.g.} \cite{cosner12,metanl}). 

In order to describe the use of the latter neighborhood we introduce the discrete convolution operator of two vectors $\mathbf x$ and $\mathbf y$ and denote it by $\mathbf x*\mathbf y.$  The definition of dicrete convolution is given as follows: 
\be (\mathbf x*\mathbf y)_q=\sum_{m\in S} x_qy_{<q-m>_Q}\ee 
where ${<q-m>_Q}=q-m ~\text{modulo}~ Q. $ Hence, the average frequency of type $A$ individuals $\mathcal C_1^q(\mathbf x)$ around any site $q\in S$ can be written as a discrete convolution operator 
\be \mathcal C_1^q(\mathbf x)=(\mathbf c^{(1)}*\mathbf x)_q=\sum_{r\in S} c^{(1)}_{<q-r>_Q}x_r\ee 
where $\mathbf c^{(1)}$ is the interaction kernel whose nonzero elements are given as $c^{(1)}_0=c^{(1)}_1=c^{(1)}_{Q-1}=\frac{1}{3}.$ Hence \eqref{general} with an appropriate time scaling can be written as
 \be\label{1D}\dot x_q=\delta(\Delta_1\mathbf x)_q+x_q(1-x_q)\big(\alpha(\mathbf c^{(1)}*\mathbf x)_q-\beta \big)\ee 
where parameters $\alpha$ and $\beta$ are as defined in \eqref{ab}.

We linearize \eqref{1D} by using the first order expansion $\mathbf{x}=\mathbf e+\varepsilon \check{\mathbf x}e^{\lambda t},$ where $\check{\mathbf x}$ is a vector independent of time $t$ denoting a spatial perturbation term. Hence this formula can be considered as a separation of variables formula for \eqref{1D}. Substituting this ansatz to the equation gives the following first order relation:

\be\label{tdfs} \lambda^{(1)} \tilde x_q=\delta(\Delta_ 1\mathbf{\tilde x})_q+\alpha e(1-e)(\mathbf c^{(1)}*\mathbf x)_q\ee

To obtain an explicit expression for the eigenvalues of the above given linear equation we need to use discrete Fourier series (DFS). Following \cite{dft,mandal}, DFS of a vector $\mathbf x$ is denoted by $\mathcal F\mathbf x$ and given by \be (\mathcal F\mathbf x)_k =\sum_{n\in S}x_n e^{{-2j\pi kq}/{Q}}\ee for any $k\in S.$ Note that $j$ is used to denote the imaginary unit number.

One can easily see that the DFS of the convolution of two vectors $\mathbf x$ and $\mathbf y$ is given by the Hadamart product of DFS of these vectors {\it i.e.} we have:
 \be (\mathcal F{\mathbf x*\mathbf y})_k=(\mathcal F\mathbf x)_k(\mathcal F\mathbf y)_k. \ee 
For a rewiev of these results, see \cite{dft,mandal}.

Note that the operator $\Delta_1$ can be written as the sum of a convolution operator and an identity operator (see \ref{ap1}). Hence taking the DFS of both sides of equation \eqref{tdfs} one obtains explicit equalities for the eigenvalues $\lambda^{(1)}_k$ as follows:
\be\label{eigen1} \lambda_k^{(1)}=\delta\Big(\omega_k^1-1\Big)+\frac{\alpha}{3} e(1-e)\Big(1+2\omega_k^1\Big)\ee
where $w_k^1=\cos\Big({\frac{2\pi k}{Q}}\Big)$ for any $k\in S.$ The explicit calculations are given in \ref{ap1}. In particular, we obtained the above equality by using formulas \eqref{dfsd} and \eqref{dfsk}.

\subsection{Calculating Eigenvalues in 2D Habitat}\label{eig2}

Extending the 1D neighborhood to 2D may lead us to different neighborhoods. We consider two major neighborhoods used both in game theory literature for cellular automata simulations (see {\it e.g.} \cite{nowak}) and in metapopulation studies (see {\it e.g.} \cite{doebelli}). First consider the von Neumann neighborhood of site $q=(q_1,q_2)\in S=\{0,1,\cdots,Q_1-1\}\times \{0,1,\cdots,Q_2-1\}$ which can be defined as folows:
\be\mathcal N_N^{q}=\{(r_1,r_2): \big<|r_1-q_1|\big>_{Q_1}+\big<|r_2-q_2|\big>_{Q_2}\leq1\}.\ee 
Another extension of the one dimensional nearest neighbour interaction range is so called Moore neighborhood defined as follows: 
\be \mathcal N_M^{q}=\{(r_1,r_2): \big<|r_1-q_1|\big>_{Q_1}\leq1,\big<|r_2-q_2|\big>_{Q_2}\leq1 \}.\ee 
Hence dispersal neighborhoods are given as $N_n^{q}=\mathcal N_n^{q}-\{q\}$ for $n=N,M.$ Note that these two neighborhood types are the simplest generalizations of 1D case. Projecting both neighborhoods to 1D leads us to the one dimensional neighborhood $N_1^q.$

Now we are ready to define the discrete dispersal operators for von Neumann and Moore neighborhoods as follows:

\be (\Delta_N \mathbf x)_q= \frac{1}{4}\sum_{r\in N_n^{q}} x_r -x_q \text{ ~~ and ~~} (\Delta_M\mathbf x)_q= \frac{1}{8}\sum_{r\in N_n^{q}} x_r -x_q.\ee

.

To write down the CRE in 2D habitat, we need to define the 2D discrete convolution operator of two vectors (or matrices) $\mathbf x$ and $\mathbf y$ which is given as follows: 
\be (\mathbf x*\mathbf y)_k= \sum_{m=0}^{Q_1}\sum_{n=0}^{Q_2} x_{q_1,q_2}y_{<m-k_1>_{Q_1},<n-k_2>_{Q_2}}\ee 
Therefore, \eqref{general} in 2D habitat $S$ with an appropriate time scaling can be written as follows:

\be\label{2D}\dot x_q= {\delta}(\Delta_n \mathbf x)_q+x_q(1-x_q)\big(\alpha(\mathbf c^{(n)}*\mathbf x)_q-\beta\big)\ee 
for $n=N,M.$ Here $\mathbf c^N$ and $\mathbf c^M$ are used to denote discrete uniform probabiliy distributions on the neighborhoods $\mathcal N_N^q$ and $\mathcal N_M^q$ of any site $q,$ respectively. The explicit expressions for these discrete probability distributions are given in \ref{ap2}.

Following \cite{dft,mandal}, discrete Fourier series (DFS) of a matix $\mathbf x=(x_{m,n})_{(m,n)\in S}$ is denoted by $\mathcal F\mathbf x$ and given by
 \be (\mathcal F\mathbf x)_k =\sum_{(m,n)\in S}x_{m,n} e^{{-2j\pi k_1n}/{Q_1}} e^{{-2j\pi k_2m}/{Q_2}}\ee 
for any $k=(k_1,k_2)\in S.$ Note that the discrete convolution theorem also holds in 2D and states that 
\be (\mathcal F \mathbf x*\mathbf y)_k=(\mathcal F \mathbf x)_k\cdot (\mathcal F \mathbf y)_k.\ee

Similarly using the first order expansion $\mathbf{x}=\mathbf e+\varepsilon \check{\mathbf x}e^{\lambda t},$ and taking the DFS of resulting linear equation, we obtain the explicit expressions for the eigenvalues.

Using equalities \eqref{dfsN} and \eqref{dfskN}, one can obtain the eigenvalues of the linearization of CRE with von Neumann neighborhood as follows:
\be\label{neumann} \lambda_k^{(N)}=\delta\Big(\frac{1}{2}\omega_k^N-1\Big)+\frac{\alpha}{5} e(1-e)\Big(1+4\omega_k^N\Big)\ee 
where $\omega_k^N=\cos\Big({\frac{2\pi k_1}{Q_1}}\Big) +\cos\Big({\frac{2\pi k_2}{Q_2}}\Big)$ for $k=(k_1,k_2)\in S.$

Similarly we obtain the eigenvalues of the linear equation with Moore neighborhood by using \eqref{dfsM} and \eqref{dfskM} as follows:
\be\label{moore} \lambda_k^{(M)}=\delta\Big(\frac{1}{4}\omega_k^M-1\Big)+\frac{\alpha}{9} e(1-e)\Big(1+8\omega_k^M\Big)\ee 
where $\omega_k^M=\cos\Big({\frac{2\pi k_1}{Q_1}}\Big) +\cos\Big({\frac{2\pi k_2}{Q_2}}\Big)+\cos\Big(2\pi \big(\frac{k_1}{Q_1}-\frac{k_2}{Q_2}\big)\Big)+\cos\Big(2\pi\big(\frac{k_1}{Q_1}+\frac{k_2}{Q_2}\big)\Big).$ The details of the derivation of these formulae are provided in \ref{ap2}.

\subsection{Generalization of the CRE and instability of the ESS}

In the above subsections we only considered the nearest neighbor interactions and derived the explicit  expressions for the eigenvalues of the linearization of CRE. It is possible to write this equation down in a more general form and identify the conditions on the interaction kernel and dispersal rate to destabilize the ESS. 
Recall the discussion in Section \ref{eig1} stating that the discrete dispersal operator $\Delta$ can be written as the sum of a convolution and the identity operator {\it i.e.} $\Delta\mathbf x=\mathbf d*\mathbf x-\mathbf x$ for a discrete probability distribution $\mathbf d.$ This is also true in 2 spatial dimensions (see \ref{ap2}). Hence we can write the CRE as \be\label{gen}\dot x_q= {\delta} \big( (\mathbf d*\mathbf x)_q-x_q\big)+x_q(1-x_q)\big(\alpha(\mathbf c*\mathbf x)_q-\beta\big)\ee 
where $\mathbf c$ and $\mathbf d$ are some discrete probability kernels. Linearizing this equation around $\mathbf e$ as done in previous subsections and taking the DFS leads us to the following eigenvalues:
 \be\lambda_k=\delta\big((\mathcal F\mathbf d)_k-1\big)+\alpha e(1-e) (\mathcal F\mathbf c)_k.\ee 
Since both $\mathbf d$ and $\mathbf c$ are discrete probability distributions their DFS (which is fully determined by the characteristic functions of the kernels) takes values  inbetween $-1$ and $1.$ Hence the real part of the term $(\mathcal F\mathbf d)_k-1$ must be non-positive. This implies that the dispersal has a stabilizing effect on the ESS. Destabilization of the ESS is only possible if the real part of the term $(\mathcal F\mathbf c)_k$ is negative for some $k\in S$ and the dispersal rate is small enough. We state this result in the following theorem.
\begin{theorem}\label{thu}
The ESS of the game $\mathbf e$ is unstable under the equation \eqref{gen} if the real part of  the DFS of the interaction kernel takes negative values and the dispersal rate $\delta$ is small enough.
\end{theorem}

Nonlocal interactions for a given number of groups result in the overlapping of foraging areas that can be thaught as a mechanism altering the frequencies of morphs
forming clusters.  The above result, therefore, may imply the existence of clusters altering around the ESS for some foraging range and frequencies determined by the nonlocal interaction kernel $\mathbf c.$

In Section \ref{intro} we mentioned the studies of evolutionary games via replicator-diffusion equations. With the assumption that all diffusion rates are equal; replicator-diffusion equations describe a monotone operator and one cannot seek a mechanism  destabilizing the ESS of the game. To study diffusion driven intabilities \cite{vickers,vickers2} introduced a special form of the population regulation to
allow for different diffusion rates. In particular, they used a nonlinear term providing a local
regulation of the frequencies of the morphs to keep solutions to replicator-diffusion equations in an extension of the game simplex. The following result related to generalized CRE ensures that the frequency of each type stays in between 0 and 1 at each patch {\it i.e.} it stays in the extended simplex $[0,1]^{|S|}$ where $|S|$ is the number of sites into which habitat is divided. 

\begin{theorem}\label{th1}
For any initial datum $\mathbf x(0)\in [0,1]^{|S|},$ solution to CRE \eqref{general} stays in the extended simplex. 
\end{theorem}The proof of this result is given in \ref{aprof}. Note also that the above theorem is valid not only for the Hawk-Dove game but all symmetric two player games. 

Now we verify that the hyphotheses of the Theorem \ref{thu} holds for the 1D and 2D cases examined above. To shorten the analysis we write the eigenvalues in a common form. Recall that the cardinality of the neighborhood $N^n$ is denoted by $s^n$ for $i=1,N,M.$ Hence all three eigenvalues can be written as follows:
 \be \lambda_k^{(n)} =\delta \Big(\frac{2}{s^n}\omega_k^n-1\Big)+\frac{\alpha}{s^n+1}e(1-e)(1+s^n\omega_k^n) \ee 
for $n=1,N,M.$ Note that \be -\frac{s^n}{2}\leq\omega_k^n\leq\frac{s^n}{2}.\ee This implies that $\frac{2}{s^i}\omega_k^i-1<1$ for all $k\neq0.$ Hence we can conclude that the dispersal has a stabilizing effect on the ESS for any $\delta >0.$

\begin{remark}\label{remark}
Recall the weak dispersal hypothesis mentioned in Section \ref{mod}. If this hypothesis is removed then \eqref{general} can be written as \beq\dot x_q=\mu(\Delta \mathbf x)_q+w f_q(\mathbf x, \mathbf c*\mathbf x)\eeq for a function $\mathbf f=(f_q)_{q\in S}.$  From the above discussion, it is clear that the eigenvalues of the linearization of this equation are given by $\mu \Big(\frac{2}{s^n}\omega_k^n-1\Big)+O(w)$ and hence the ESS of the game is stable.
\end{remark}

Since all three nearest neighbour interaction kernels are related to the uniform distribution, the DFS of each kernel  take negative values. The DFS of the kernel $\mathbf c^{(1)},$ for example, takes negative values since $1+2\cos\Big({\frac{2\pi k}{Q}}\Big)<0$ for some values of $k$ near $Q/2.$ Similarly, it can be easily seen that the DFS of the interaction kernels take negative values for von Neumann and Moore neighborhoods.

Suppose that $\delta=0$ then the ESS of the game is unstable for nearest neighbour interactions in 1D or 2D habitats. By continuity it is also unstable for $\delta$ small enough. On the other hand it is stable for large enough $\delta>0.$ Hence, we conclude that there exists a critical diffusion rate $\delta_0$ such that the ESS is stable for $\delta>\delta_0$ and unstable for $\delta<\delta_0.$ Such a critical diffusion rate can be calculated by solving the following root finding problem:
 \be\label{cdr} \max_{k\in S} \lambda_k^{(n)}(\delta) =0\ee for $n=1,N,M.$  The argument of this maximization problem is denoted by $k_c$ and called the most unstable wavenumber.

\section{Numerical Simulations}
\label{ns}
Throughout this section, we consider the following payoff matrix \[\begin{tabular}{ l | c r }
& A & B \\ \hline
A & $3$ & $ 1$ \\ 
B & $4$ & $ 0$ \\
\end{tabular}
\] descrribing a Hawk-Dove game with $\alpha=-2,$ $\beta=-1$ and $e=1/2.$

First we consider the equation \eqref{1D} in a 1D habitat forming a ring. Our model with nearest neighborhood interaction and dispersal gives rise to two types of patterns in phenotype space.  The emergence of these two types of patterns are related to the number of patches contained in the habitat $S.$ In particular the number of patches determines the number of competing eigenvalues and the emergence of different spatial patterns.

\begin{figure}[h!]

\centering
\subfigure[]{
\includegraphics[width=7cm, height=7cm]{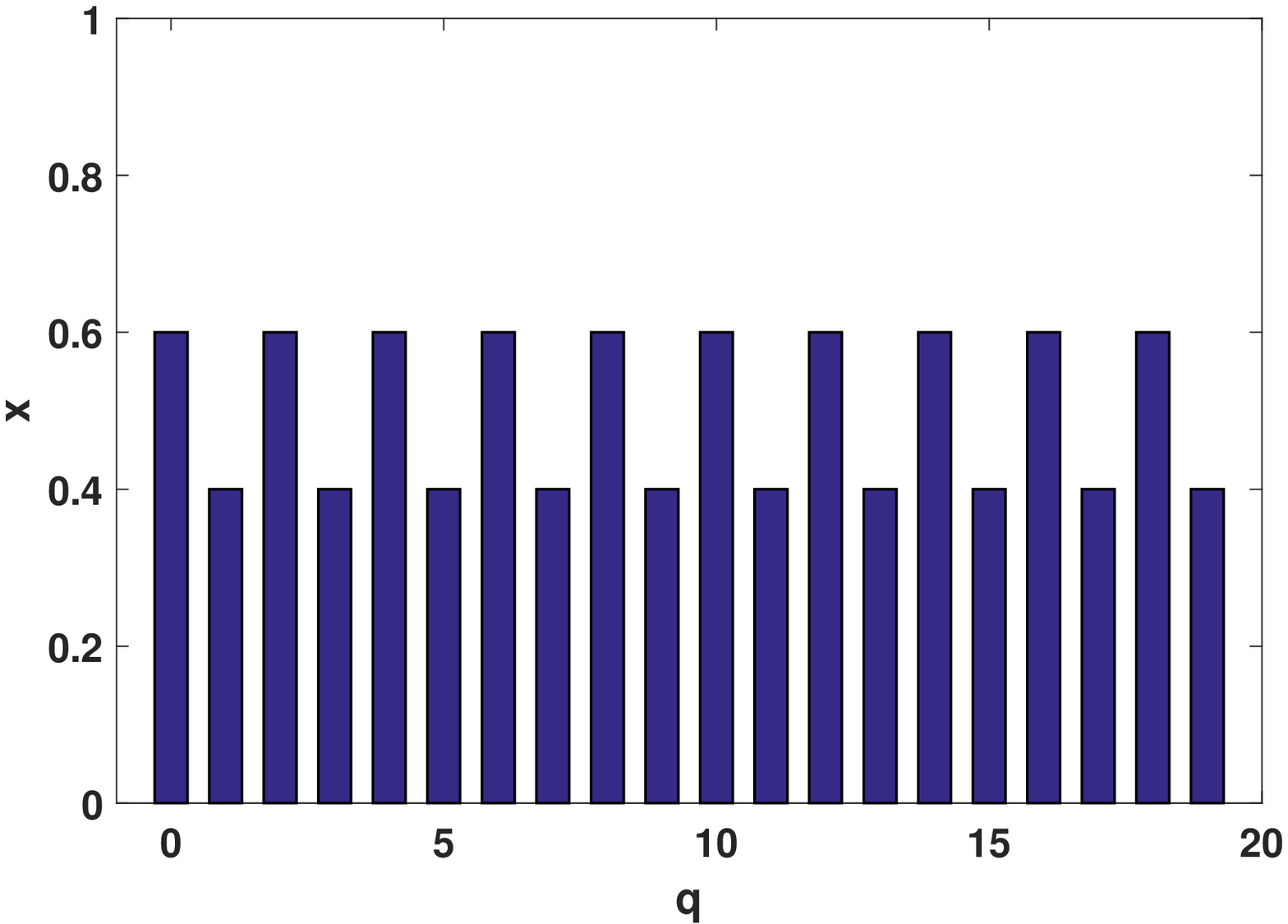}
}
\subfigure[]{
\includegraphics[width=7cm, height=7cm]{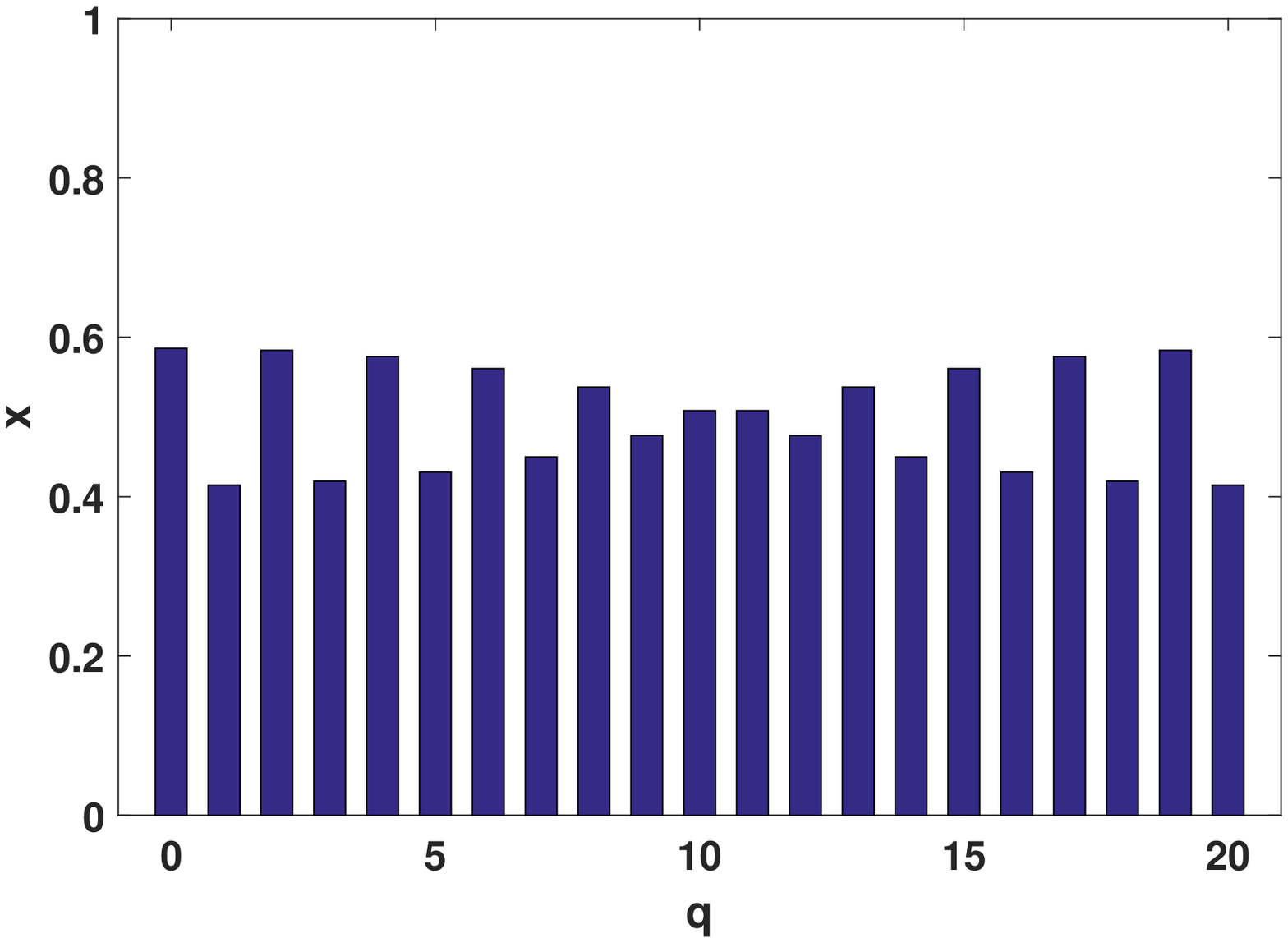}
} 
\caption{Stationary waves for the dispersal rate $\delta=0.08$. Panel (a) illustrates regular stationary waves for the payoff matrix given above and $Q=20$. Similarly, Panel (b) shows degenerate stationary waves for thesame payoff matrix and $Q=21$. }
\label{1d}
\end{figure}

Figure \ref{1d} panels (a) and (b) illustrates these patterns for $Q=20$ and $Q=21.$ For $Q=20$ there exists only one most unstable wave number giving rise to patterns shown in Figure \ref{1d} Panel (a). More generally, when the number of patches are even we have only one the most unstable wavenumber which is given by $k_c=Q/2$ on which the DFS of the interaction kernel takes its  smallest value. We  refer to these patterns as {\it regular} stationary-wave type patterns. On the other hand, when the patch number is odd there are two competing wavenumbers causing the emergence of a different type of stationary patterns. An example of such patterns for $Q=21$ is shown in Figure \ref{1d} Panel (b). We  refer to these patterns as {\it degenerate} stationary patterns.
For our numerical calculations we used \eqref{cdr} and the eigenvalues \eqref{eigen1} to calculate the critical dispersal rates for the parameters given in the above payoff matrix. Having this value enables us to calculate the most unstable wave number(s) giving rise to patterns shown in Figure \ref{1d}. The critical dispersal rates and associated wave numbers for $Q=20$ and $Q=21$ are given as follows.
\begin{itemize}
\item $\delta_0=0.0833$ for $Q=20$ for which the most unstable wave number is associated with the eigenvalue $\lambda_{10}^{(1)}$ 
\item $\delta_0=0.0819$ for $Q=21$ where we get two competing eigenvalues namely $\lambda^{(1)}_{10}$ and $\lambda^{(1)}_{11}.$
\end{itemize}

\begin{figure}[h!]

\centering
\begin{subfigure}[]{}
\includegraphics[width=7cm, height=7cm ]{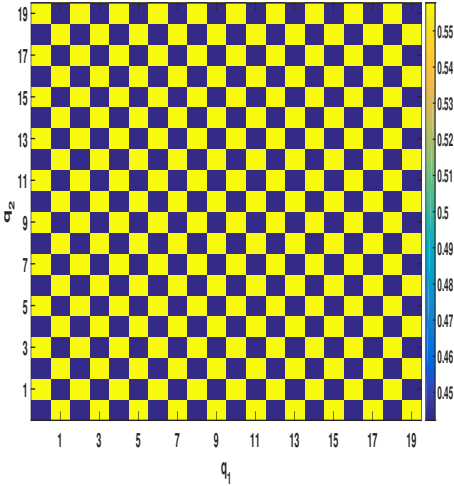}
\end{subfigure}

\begin{subfigure}[]{}
\includegraphics[width=7cm, height=7cm]{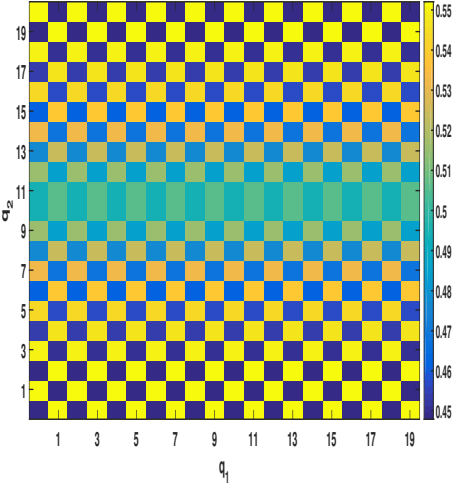}
\end{subfigure}
\begin{subfigure}[]{}
\includegraphics[width=7cm, height=7cm]{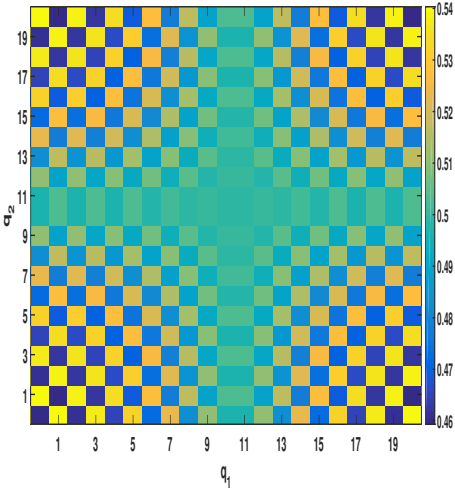}
\end{subfigure}
\caption{Patterns arising from 2D CRE with von Neumann neighborhood and dispersal rate $\delta=0.148.$ Panel (a) illustrates regular stationary waves in 2D domain with $Q_1=Q_2=20.$ Panel (b) shows semi-degenerate stationary waves in 2D with $Q_1=20$ and $Q_2=21$ for the same payoff matrix. Panel (c) illustrates the degenerate stationary waves in 2D with $Q_1=Q_2=21.$ }
\label{2n}
\end{figure}

Next we consider the equation \eqref{2D} in a 2D habitat with von Neumann neighborhood. Again the types of patterns arising from this equation  are related to the number of patches contained by the habitat. For $q=(q_1,q_2)\in S$ we observed regular stationary patterns (see Figure \ref{2n} Panel (a)) for $Q_1=Q_2=20$ for which we have only one unstable wavenumber. For $Q_1=20$ and $Q_2=21,$ we observed the emergence of patterns shown in Figure \ref{2n} Panel (b). We would like to call attention to a similarity between these patterns and the ones shown in Figure \ref{1d}. For a fixed $q_1,$ one can easily see that these patterns are similar to the one obtained in Figure \ref{1d} Panel (b). On the other hand, the obtained patterns are similar to 1D patterns shown in \ref{1d} Panel (a) for fixed $q_2.$ Hence we call these patterns {\it semi-degenerate} stationary patterns.  Finally, considering the case $Q_1=Q_2=21$ leads us to the patterns shown in Figure \ref{2n} Panel (c). These patterns can be thought as a 2D degenetarte stationary patterns since fixing either $q_1$ or $q_2$ results in degenerate 1D degenerate patterns shown in Figure \ref{1d} Panel (b).

For our numerical calculations we again used \eqref{cdr} and the eigenvalues \eqref{neumann} to calculate the critical dispersal rates for the game parameters given in the above payoff matrix. The critical dispersal rates and associated wave numbers for different values of $Q_1$ and $Q_2$ are given as follows.

\begin{itemize}
\item $\delta_0=0.15$ for $Q_1=Q_2=20.$ Here the most unstable wavenumber is associated with the eigenvalue $\lambda^{(N)}_{(10,10)}.$
\item $\delta_0=0.1493$ for $Q_1=20$ and $Q_2=21.$ Here we get two competing eigenvalues namely $\lambda^{(N)}_{(10,10)}$ and $\lambda^{(N)}_{(10,11)}.$

\item $\delta_0=0.1486$ for $Q_1=Q_2=21.$ Here we get four competing eigenvalues namely $\lambda^{(N)}_{(10,10)},$ $\lambda^{(N)}_{(10,11)},$ $\lambda^{(N)}_{(11,10)}$ and $\lambda^{(N)}_{(11,11)}.$
\end{itemize} 
We remark that the values of $Q_1$ and $Q_2$ are chosen only for illustration purposes. The factor determining the number of unstable wavenumbers is related to the evenness or oddness of these values as in the 1D case.

\begin{figure}[h!]

\centering
\begin{subfigure}[]{}
\includegraphics[width=7cm, height=7cm ]{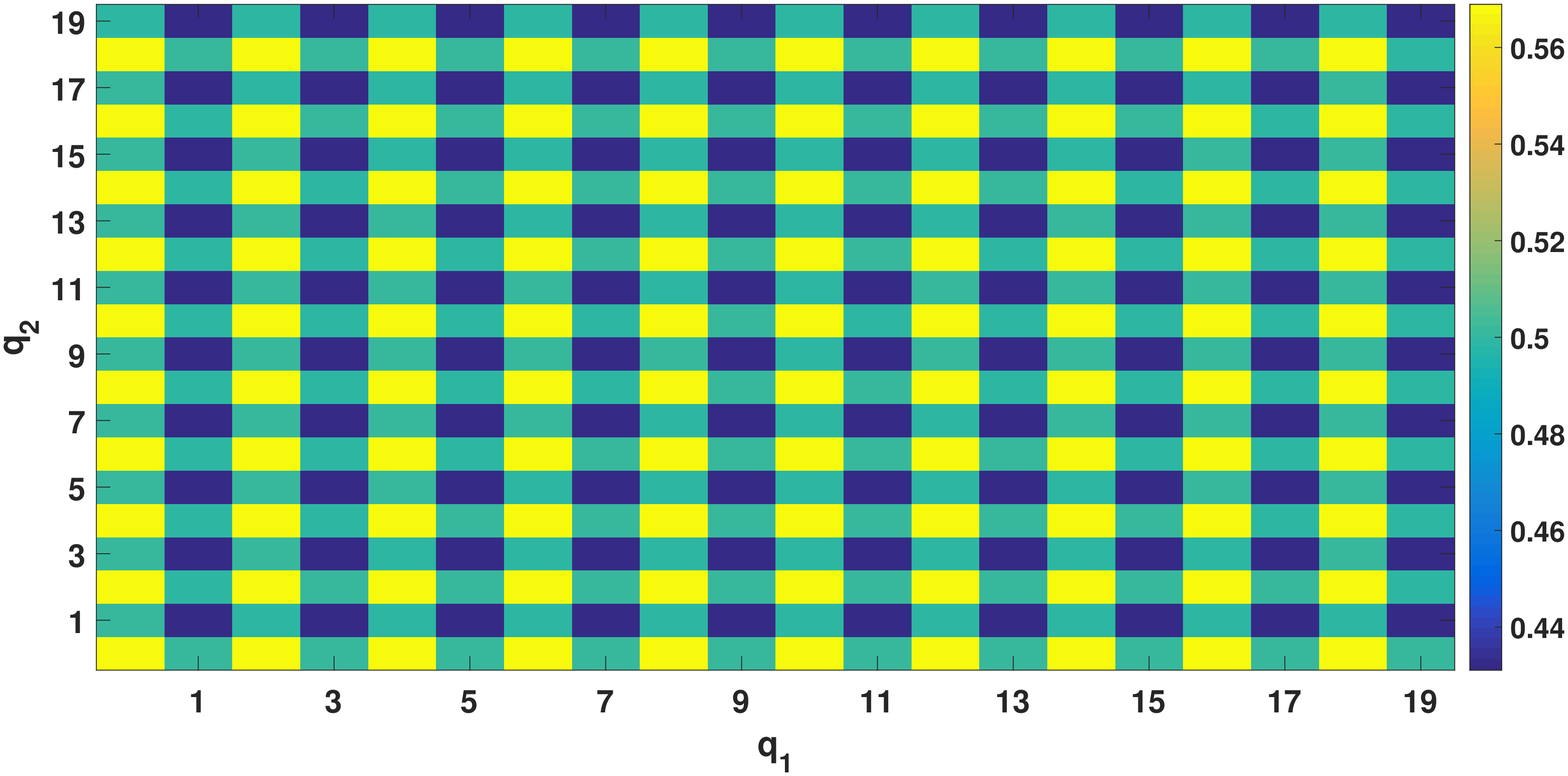}
\end{subfigure}

\begin{subfigure}[]{}
\includegraphics[width=7cm, height=7cm]{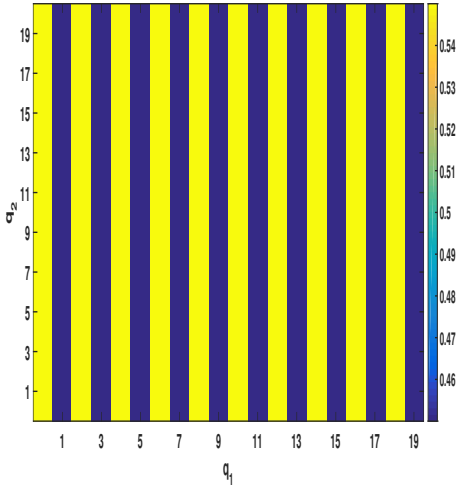}
\end{subfigure}
\begin{subfigure}[]{}
\includegraphics[width=7cm, height=7cm]{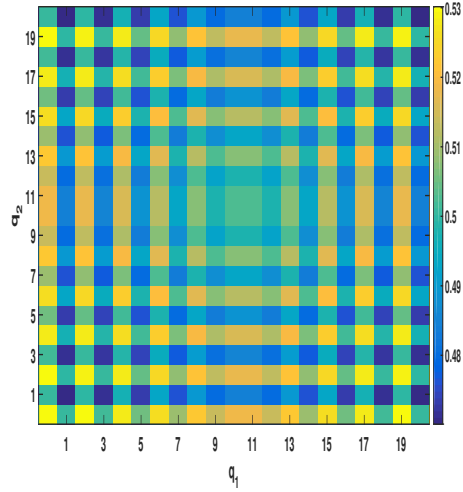}
\end{subfigure}

\caption{Patterns arising from 2D CRE with Moore neighborhood for the payoff matrix defined above. Panel (a) illustrates regular stationary waves in 2D domain with $Q_1=Q_2=20$ for dispersal rate $\delta=0.0109.$ Panel (b) shows stripe type patterns in 2D with $Q_1=20$ and $Q_2=21$ for $\delta=0.11.$ Panel (c) illustrates a different kind of degenerate stationary waves in 2D with $Q_1=Q-2=21$ for $\delta=0.0109.$ }
\label{2m}
\end{figure}

Finally we consider the equation \eqref{general} in 2D habitat with Moore neighborhood. Again the patterns arising from this equation is related to the number of patches contained by the environment. We observed patterns shown in Figure \ref{2m} Panel (a) for $Q_1=Q_2=20$ for which there exist two competing unstable wavenumbers. This patters can be considered as a composition of two different 1D regular patterns. If the patch numbers are taken as $Q_1=20,$ $Q_2=21$ we observe strip patterns shown in Figure \ref{2m} Panel (b). For a fixed $q_2,$ one can easily see that these patterns are similar to the regular patterns obtained in Figure \ref{1d} Panel (a). When $Q_1=Q_2=21$ we observe the patterns shown in Figure \ref{2n} Panel (c). These patterns can be considered as a composition of two different 1D degenetarte patterns.

For our numerical calculations we again used \eqref{cdr} and the eigenvalues \eqref{moore} to calculate the critical dispersal rates for the parameters given by the above payoff matrix. The critical dispersal rates and associated wave numbers for different values of $N$ and $M$ are given as follows.

\begin{itemize}
\item $\delta_0=0.1111$ for $Q_1=Q_2=20$ Here we get two competing eigenvalues namely $\lambda^{(M)}_{(10,20)}$ and $\lambda^{(M)}_{(20,10)}.$

\item $\delta_0=0.1111$ for $Q_1=20$ and $q_2=21.$ Here the most unstable wavenumber is associated with the eigenvalue $\lambda^{(M)}_{(10,21)}.$

\item $\delta_0=0.1092$ for $Q_1=Q_2=21$ Here we get four competing eigenvalues namely $\lambda^{(M)}_{(10,21)},$ $\lambda^{(M)}_{(11,21)},$ $\lambda^{(M)}_{(21,10)}$ and $\lambda^{(M)}_{(21,11)}.$
\end{itemize}

It is easy to observe that Moore neighborhood gives rise to more complex patterns in general. The reason behind this observation is that the eigenvalues associated with Moore neighborhood identifies distinct wavenumbers. Whereas all of the most unstable wavenumbers asociated with the von Neumann neighborhood are near $(Q_1/2,Q_2/2).$

\section{Discussion}\label{discussion}

In this paper, we have obtained the coupled replicator equations (CRE) for two player symmetric games as mean dynamics of a frequency dependent Moran process coupled via spatial dispersal and nonlocal payoff calculation in a patchy environment. In particular, we studied the stability of the mixed strategy ESS of the Hawk-Dove game in this patchy  environment. We have shown that the spatial dispersal operator has a stabilizing effect on the spatially homogenous solution of CRE ({\it i.e.} mixed strategy ESS of the game) as in other ecological models (see for example \cite{hastings1,mig,doebeli2}) while nonlocal payoff calculation may destablize the ESS. In  this paper we only explicitly calculate the Fourier transforms for kernels modeling nearest neighborhood interactions and dispersal in 1D and 2D periodic habitats to keep the technicality to a minimum. However, we also explicitely stated the form that the CDE takes when one  needs to consider more complex dispersal and interaction structures to incorporate movement to further patches, or to include unequal weighting. As a result, we found that the ESS of the game is unstable if the (real part of) interaction kernel takes negative values and the dispersal rate is sufficiently small. 

We also numerically studied the patterns arising from CRE with nearest neighbour symmetric dispersal and interaction kernels. We observed that the evenness or oddness of the number of sites affect the shape of the spatial patterns. If the number of patches are even and the habitat is one dimensional, the arising patterns are similar to the regular stationary wave type patterns arising from reaction diffusion equations. Whereas, If the number of patches are odd we observed the formation of degenerate stationary patterns (see Figure \ref{1d} Panel (b)) For von Neumann neighborhood, we see that the patterns in 2D habitat as shown in Figure \ref{2n} are just 2D generalizations of the ones observed in Figure \ref{1d}. On the other hand, our numerical study indicates that Moore neighbourhood gives rise to the emergence of more complex patterns when compared to the ones arising from CRE coupled via von Neumann neighborhood (see Figure \ref{2m}).

\begin{figure}[h!]

\centering
\includegraphics[width=8.25cm, height=8.5cm]{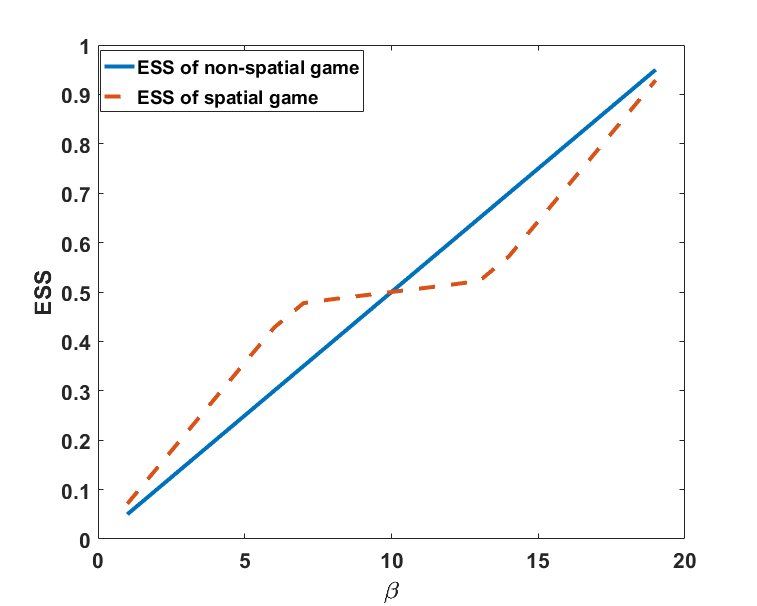}
\caption{Comparison of ESS with the spatial average of frequencies of type A individuals for $\alpha=-20$ and a very small dispersal rate $\delta=0.0001.$ The blue line is the ESS of the nonspatial game and the red curve is the dynamically stable average frequencies in 1D patchy environment. }
\label{comp}

\end{figure}

The effect of spatial structure in discrete individual based models was first mentioned by \cite{NM,NM1} for the Hawk-Dove game. Subsequently, it was discussed in more detail by \cite{KD} whose numerical simulations indicated that, the consideration of the spatial structure enables doves to become spontaneously organised into
clusters so that they can outweigh losses against hawks. The patterns obtained here can be interpreted as a mechanism for spontaneous
cluster formation as done by \cite{hauertsp}. Hence our results  support the numerical  findings of \cite{KD}.  However, \cite{hauert02} considered the Hawk-Dove game on a lattice and remarked that spatial extension generally favours the hawk strategy when compared to the replicator equations. Our model, however, suggests that a strategy is favored when compared to the mean field dynamics only if the ESS corresponding to this strategy is less than $1/2$ (see Figure \ref{comp}).

It was  revealed in the literature that pattern formation can be observed when modified replicator-diffusion models are used \cite{vickers}. In addition, there are  further studies exploring spatially extended public goods games and it was shown that reaction-diffusion equations modeling the public good phenomenon exhibits not only pattern formation but also chaotic dynamics \cite{hauertsp}. In both continuous space models it was assumed that the diffusion rates of two morphs are different. Thus instabilities observed from these models are  essentially diffusion-driven instabilities. Contrary to these studies our model takes nonlocal interactios or payoff calculation into account. Such interactions can be considered as another mechanism destabilizing the ESS. 

Finally, there are several noteworthy issues that should be further explored or extended. In this paper, we focused on two-strategy games with an interior ESS. The derived results may be extended to $n$-strategy games with $n\geq 3$, such as the Rock-Scissors-Paper game. In addition, following \cite{aydogmus3}, a nonlinear analysis of the CRE can be performed to study the phase transitions near stability boundaries. We also would like to note that one can extend the analysis performed in Section \ref{lsa} from the spatial lattices to more general networks. Such an extension, however,  requires a well-defined Fourier-like transformation(s) on the generalized network structure. An apparent example of such graphs is Cayley graphs on which Fourier transform is defined \cite{terras}. For wavelet transformation on more general graphs, see \cite[Section 8.2]{image}.

\appendix

\section{Explicit Calculation of Eigenvalues}
\subsection{1D Case}
\label{ap1}

Note that the discrete dispersal operator can be written as $\Delta_1=\mathbf d^1 -I^1$ where $\mathbf d^1=(d_n^1)_{n\in S}$ is the discrete probability kernel on $S$ with $d_1^1=d_{Q-1}^1=1/2$ and $I$ is the identitiy operator. By definition of one dimensional DFS one can calculate the DFS of the kernel $\mathbf d^1$ as follows:

\bea (\mathcal F\mathbf d^1)_k &=&\sum_{q\in S}d_n^1 e^{{-2j\pi kq}/{Q}}\nonumber\\
&=&\frac{1}{2}(e^{{-2j\pi k}/{Q}}+e^{{-2j\pi k(Q-1)}/{Q}})\nonumber\\
\label{dfsd}&=&\cos\Big({\frac{2\pi k}{Q}}\Big)
\eea

Similarly one can calculate the DFS of the interaction kernel $\mathbf c^1$ with $c_0=c_1=c_{N-1}=1/3$ as \be\label{dfsk}(\mathcal F\mathbf c^1)_k=\frac{1}{3}+\frac{2}{3}\cos\Big({\frac{2\pi k}{Q}}\Big)\ee

\subsection{2D Cases}
\label{ap2}
 
Suppose that if the individuals interact in a von Neumann neighborhood, the discrete dispersal operator can be written as $\Delta_N=\mathbf d^N -I$ where $\mathbf d^N$ is the discrete probability kernel on $S$ with $d_{1,0}^N= d_{0,1}^N=d_{Q_1-1,0}^N=d_{0,Q_2-1}^N=1/4$ and $I$ is the identitiy operator. By definition of two dimensional DFS one can calculate the DFS of the kernel $\mathbf d^N$ as follows:

\bea (\mathcal F\mathbf d^N)_k &=&\sum_{(q_1,q_2)\in S}d_{q_1,q_2}^N e^{{-2j\pi kq_1}/{Q_1}} e^{{-2j\pi kq_2}/{Q_2}}\nonumber\\
&=&\frac{1}{4}(e^{{-2j\pi k_1}/{Q_1}}+e^{{-2j\pi k_2}/{Q_2}}+e^{{-2j\pi k_1(N-1)}/{Q_1}}+e^{{-2j\pi k_2(N-1)}/{Q_2}})\nonumber\\
\label{dfsN}&=&\frac{1}{2}\Big(\cos\Big({\frac{2\pi k_1}{Q_1}}\Big) +\cos\Big({\frac{2\pi k_2}{Q_2}}\Big)\Big)
\eea
The corresponding interaction kernel is $\mathbf c^N$ given by $c_{1,0}=c_{0,1}=c_{N-1,0}=c_{0,N-1}=c_(0,0)=1/5$ and its DFS is given as follows:
 
\be\label{dfskN}(\mathcal F\mathbf c^N)_k=\frac{1}{5}+\frac{2}{5}\Big(\cos\Big({\frac{2\pi k_1}{Q_1}}\Big) +\cos\Big({\frac{2\pi k_2}{Q_2}}\Big)\Big)\ee

Assuming Moore neighborhood, we get the following equalities:
\be
\label{dfsM}
(\mathcal F\mathbf d^M)_k =\frac{1}{4}\Bigg(\cos\Big({\frac{2\pi k_1}{Q_1}}\Big) +\cos\Big({\frac{2\pi k_2}{Q_2}}\Big)+\cos\Big(2\pi \big(\frac{k_1}{Q_1}-\frac{k_2}{Q_2}\big)\Big)+\cos\Big(2\pi\big(\frac{k_1}{Q_1}+\frac{k_2}{Q_2}\big)\Big)\Bigg)\ee and \be \label{dfskM}(\mathcal F\mathbf c^M)_k=\frac{1}{9}+\frac{2}{9}\Bigg(\cos\Big({\frac{2\pi k_1}{Q_1}}\Big) +\cos\Big({\frac{2\pi k_2}{Q_2}}\Big)+\cos\Big(2\pi \big(\frac{k_1}{Q_1}-\frac{k_2}{Q_2}\big)\Big)+\cos\Big(2\pi\big(\frac{k_1}{Q_1}+\frac{k_2}{Q_2}\big)\Big)\Bigg)\ee

\section{Proof of Theorem \ref{th1}}\label{aprof}
 Let $\mathbf x,\mathbf y \in \mathbb R^{|S|}$ be two functions inthe space of  bounded uniformly continuous functions on $[0,T]$  equipped with the norm 
$||\mathbf x||_T=\sup_{q,t}|x_q(t)|.$ Suppose that $0\leq y_q\leq 1$ for all $q\in S$ and let $\mathbf x$ be a solution to equation
\begin{equation}
\label{repidev}
\frac{dx_q}{d t}=\mu(\mathbf d *\mathbf y)_q-x_q+wx_q(1-x_q)(\alpha(\mathbf c*\mathbf y)_q-\beta)
\end{equation} 
with the initial data $\mathbf x(0)=\mathbf x_0.$ Here the discrete probability kernel $\mathbf d$ is assumed to be a step function with support $N^q.$ 
 Since $\mathbf d$ is a discrete probability distribution, $0\leq (\mathbf d*\mathbf y)_q\leq1$ for any $q\in S.$ Notice that ${d x_q}/{d t}=(\mathbf d*\mathbf y)_q\geq 0$ if $ x_q=0$ for some $q\in S$ and some $t\in[0,T].$ The trajectory of $\mathbf x$ is therefore always beyond 0. Similarly, ${d x_q}/{d t}\leq0$ if $x_q=1$  for some $q\in S$ and some $t\in[0,T].$ Hence, $\mathbf x\in [0,1]^{|S|}$ for any $t\in [0,T].$

 { Now we will show that mapping $\mathbf y \to \mathbf x$ is a contraction. Let $\mathbf y^1$ and $\mathbf y^2$ be two functions satisfying the above conditions and suppose that $\mathbf x^1$ and $\mathbf x^2$ denote the corresponding solutions to equation \eqref{repidev} with the same initial data. Let $\mathbf x=\mathbf x^1-\mathbf x^2$ and $\mathbf y=\mathbf y^1-\mathbf y^2$, so that \begin{equation}
\label{short}
\frac{d x_q }{d t}=
\Upsilon_1^q x_q+ \mu(\mathbf d*\mathbf y)_q +\Upsilon_2^q (\mathbf c*\mathbf y)_q,
\end{equation} where $\Upsilon_1^q=w\alpha(y^1_q-y^2_q(x^1_q+x^2_q))+w\beta(-1+x^1_q+x^2_q)$ and $\Upsilon_2^q= w\alpha(x^2_q-(x^1_q)^2)$.
Integration up to time $t$ yields
\begin{eqnarray*}
x_q=\int_0^t \Upsilon_1 x_q\,ds+\int_0^t  \Upsilon_2 (\mathbf c*\mathbf y)_q +\mu (\mathbf d*\mathbf y)_q \,ds,\end{eqnarray*}
which implies \begin{eqnarray*}
|x_q|\leq\int_0^t |\Upsilon_1^q| |x_q|\,ds+\int_0^t |\Upsilon_2| |(\mathbf c*\mathbf y)_q|+\mu |(\mathbf d*\mathbf y)_q|\,ds.
\end{eqnarray*}
Let $K_k= \max_{x_q,y_q\in[0,1]} \Upsilon_k$ for $i,j,k=1,2$, and take the supremum over $q\in S$ and $t\in [0,T]$, we get
$$||\mathbf x||_T\leq \frac{T(K_2+\mu)}{1-TK_1}||\mathbf y||_T.$$
For sufficiently small $T$, $ {T(K_2+\mu)}/(1-TK_1)<1$. { Hence, for sufficiently small $T$,  $\mathbf y \to \mathbf x$ is a contraction preserving the closed subset $[0,1],$ and therefore admits a unique fixed point. Since $T$ is independent of the size of the data, this argument can be iterated from $T$ to any desired time interval.}

\section{The Mean-Field Equation in Strong Selection Regime}\label{ap3}
Here we consider the mean-field equation of the frequency dependent Moran process when the selection parameter is large. In particular, we suppose that $w=1$ yielding the original Markov chain studied by \cite{finite}. In this case, the fitness of each morph equals to its frequency dependent payoff \eqref{payof}. This implies the following equality: $$\psi^{q}(\mathcal X)=\frac{\mathcal X^{q}\pi_A^{q}}{\mathcal X^{q}\pi_A^{q}+(1-\mathcal X^{q})\pi_B^{q}}.$$ Using this probability, one can construct the Markov process to yield its mean dynamics. The mean-field equation of this process is given by the following equation: \beq \label{111} \dot x_q=\mu (\Delta \mathbf{\psi})_q +\psi^q-x_q\eeq where $\psi^q:=\psi^q(\mathbf x)$ and  $(\Delta \mathbf{\psi})_q=\frac{1}{s}\sum_{r\in N^q}\psi^r-\psi^q.$ Here we would like to note that the spatially homogenous solution loses its stability for sufficiently small dispersal probability $\mu.$ Figure \ref{compq} illustrates degenerate stationary patterns for Q = 21.

\begin{figure}[h!]

\centering
\includegraphics[width=9.5cm, height=8.25cm]{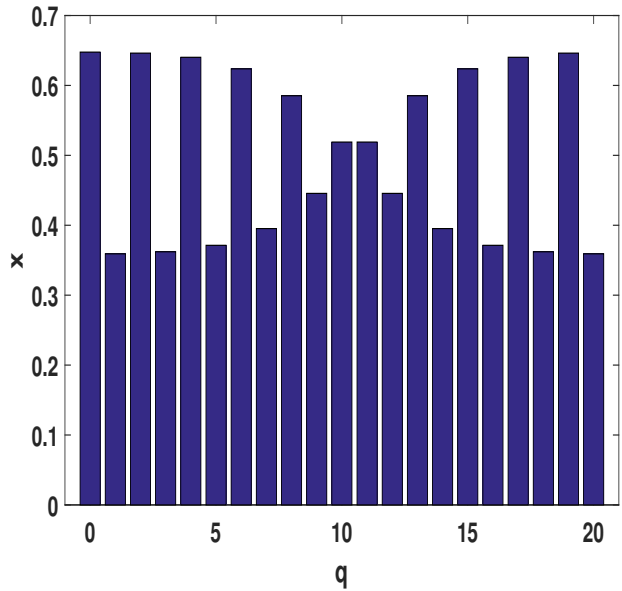}
\caption{Stationary wave type patterns  for the Hawk-Dove game whose payoffs are given in Section \ref{ns} with a dispersal probability $\mu=0.035.$  }
\label{compq}

\end{figure}




\begin{thebibliography}{100}
\expandafter\ifx\csname natexlab\endcsname\relax\def\natexlab#1{#1}\fi
\expandafter\ifx\csname url\endcsname\relax
  \def\url#1{\texttt{#1}}\fi
\expandafter\ifx\csname urlprefix\endcsname\relax\def\urlprefix{URL }\fi

\bibitem{allenowak}
Allen, B., Nowak, M.~A., 2014. Games on graphs. EMS surveys in mathematical
  sciences 1~(1), 113--151.

\bibitem{arrigoni}
Arrigoni, F., Pugliese, A., 2002. Limits of a multi-patch sis epidemic model.
  Journal of mathematical biology 45~(5), 419--440.

\bibitem{aydogmus1}
Aydogmus, O., 2015. Patterns and transitions to instability in an intraspecific
  competition model with nonlocal diffusion and interaction. Math. Mod. Nat.
  Phen. 10~(6), 17--29.

\bibitem{aydogmus3}
Aydogmus, O., 2017. Phase transitions in a logistic metapopulation model with
  nonlocal interactions. Bulletin of mathematical biology.

\bibitem{aydogmus}
Aydogmus, O., Zhou, W., Kang, Y., 2017. On the preservation of cooperation in
  two-strategy games with nonlocal interactions. Mathematical biosciences 285,
  25--42.

\bibitem{benaim}
Bena{\"\i}m, M., Weibull, J.~W., 2003. Deterministic approximation of
  stochastic evolution in games. Econometrica 71~(3), 873--903.

\bibitem{main}
Britton, N., 1989. Aggregation and the competitive exclusion principle. J.
  Theor. Biol. 136~(1), 57--66.

\bibitem{cosner12}
Cantrell, R.~S., Cosner, C., Fagan, W.~F., 2012. The implications of model
  formulation when transitioning from spatial to landscape ecology. Math.
  Biosci. and Eng. 9~(1), 27--60.

\bibitem{macro}
Chalub, F.~A., Souza, M.~O., 2009. From discrete to continuous evolution
  models: a unifying approach to drift-diffusion and replicator dynamics.
  Theoretical Population Biology 76~(4), 268--277.

\bibitem{vickers2}
Cressman, R., Vickers, G., 1997. Spatial and density effects in evolutionary
  game theory. Journal of Theoretical Biology 184~(4), 359--369.

\bibitem{doebeli2}
Doebeli, M., 1995. Dispersal and dynamics. Theor. Pop. Biol. 47~(1), 82--106.

\bibitem{doebelli}
Doebeli, M., Killingback, T., 2003. Metapopulation dynamics with quasi-local
  competition. Theor. Pop. Biol. 64~(4), 397--416.

\bibitem{FM00}
Ferriere, R., Michod, R.~E., et~al., 2000. Wave patterns in spatial games and
  the evolution of cooperation. The Geometry of Ecological Interactions:
  Simplifying Spatial Complexity, 318--339.

\bibitem{pattern}
Genieys, S., Volpert, V., Auger, P., 2006. Pattern and waves for a model in
  population dynamics with nonlocal consumption of resources. Math. Mod. Nat.
  Phen. 1~(01), 63--80.

\bibitem{mig}
Gyllenberg, M., S{\"o}derbacka, G., Ericsson, S., 1993. Does migration
  stabilize local population dynamics? analysis of a discrete metapopulation
  model. Math. Biosci. 118~(1), 25--49.

\bibitem{hastings1}
Hastings, A., 1993. Complex interactions between dispersal and dynamics:
  lessons from coupled logistic equations. Ecology 74~(5), 1362--1372.

\bibitem{hauert02}
Hauert, C., 2002. Effects of space in 2$\times$ 2 games. International Journal
  of Bifurcation and Chaos 12~(07), 1531--1548.

\bibitem{deme}
Hauert, C., Imhof, L.~A., 2012. Evolutionary games in deme structured, finite
  populations. Journal of theoretical biology 299, 106--112.

\bibitem{Hof}
Hofbauer, J., 1998. Equilibrium selection via travelling waves. In: Game
  theory, experience, rationality. Springer, pp. 245--259.

\bibitem{Hof1}
Hofbauer, J., Hutson, V., Vickers, G.~T., 1997. Travelling waves for games in
  economics and biology. Nonlinear Analysis: Theory, Methods \& Applications
  30~(2), 1235--1244.

\bibitem{HS}
Hofbauer, J., Sigmund, K., 1998. Evolutionary games and population dynamics.
  Cambridge university press.

\bibitem{ESS}
Hutson, V., Vickers, G.~T., 1992. Travelling waves and dominance of ess's.
  Journal of Mathematical Biology 30~(5), 457--471.

\bibitem{hwang}
Hwang, S.-H., Katsoulakis, M., Rey-Bellet, L., 2013. Deterministic equations
  for stochastic spatial evolutionary games. Theoretical Economics 8~(3),
  829--874.

\bibitem{KD}
Killingback, T., Doebeli, M., 1996. Spatial evolutionary game theory: Hawks and
  doves revisited. Proceedings of the Royal Society of London B: Biological
  Sciences 263~(1374), 1135--1144.

\bibitem{killing2}
Killingback, T., Loftus, G., Sundaram, B., 2013. Competitively coupled maps and
  spatial pattern formation. Physical Review E 87~(2), 022902.

\bibitem{kisdi}
Kisdi, {\'E}., Utz, M., 2005. Does quasi-local competition lead to pattern
  formation in metapopulations? an explicit resource competition model. Theor.
  Pop. Biol. 68~(2), 133--145.

\bibitem{kurtz}
Kurtz, T.~G., 1978. Strong approximation theorems for density dependent markov
  chains. Stochastic Processes and their Applications 6~(3), 223--240.

\bibitem{image}
L{\'e}zoray, O., Grady, L., 2012. Image processing and analysis with graphs:
  theory and practice. CRC Press.

\bibitem{mandal}
Mandal, M., Asif, A., 2007. Continuous and discrete time signals and systems.
  Cambridge University Press.

\bibitem{metanl}
Maruvka, Y.~E., Shnerb, N.~M., 2006. Nonlocal competition and logistic growth:
  patterns, defects, and fronts. Phys. Rev. E 73~(1), 011903.

\bibitem{smith}
Maynard~Smith, J., 1974. The theory of games and the evolution of animal
  conflicts. Journal of theoretical biology 47~(1), 209--221.

\bibitem{MSP}
Maynard~Smith, J., Price, G., 1973. The logic of animal conflict. Nature 246,
  15.

\bibitem{nowak}
Nowak, M.~A., 2006. Evolutionary dynamics. Harvard University Press.

\bibitem{NBM}
Nowak, M.~A., Bonhoeffer, S., May, R.~M., 1994. More spatial games.
  International Journal of Bifurcation and Chaos 4~(01), 33--56.

\bibitem{NM}
Nowak, M.~A., May, R.~M., 1992. Evolutionary games and spatial chaos. Nature
  359~(6398), 826--829.

\bibitem{NM1}
Nowak, M.~A., May, R.~M., 1993. The spatial dilemmas of evolution.
  International Journal of bifurcation and chaos 3~(01), 35--78.

\bibitem{NS}
Nowak, M.~A., Sigmund, K., 2000. Games on grids. The Geometry of Ecological
  Interactions: Simplifying Spatial Complexity, 135--150.

\bibitem{ohtsuki}
Ohtsuki, H., Hauert, C., Lieberman, E., Nowak, M.~A., 2006. A simple rule for
  the evolution of cooperation on graphs and social networks. Nature
  441~(7092), 502--505.

\bibitem{ohtsuki2}
Ohtsuki, H., Nowak, M.~A., 2006. Evolutionary games on cycles. Proceedings of
  the Royal Society of London B: Biological Sciences 273~(1598), 2249--2256.

\bibitem{dft}
Smith, J., 2007. Mathematics of the discrete Fourier transform (DFT): with
  audio applicaitons. W3K Publishing.

\bibitem{szabo}
Szab{\'o}, G., Fath, G., 2007. Evolutionary games on graphs. Physics reports
  446~(4), 97--216.

\bibitem{finite}
Taylor, C., Fudenberg, D., Sasaki, A., Nowak, M.~A., 2004. Evolutionary game
  dynamics in finite populations. Bulletin of mathematical biology 66~(6),
  1621--1644.

\bibitem{taylor0}
Taylor, P.~D., Jonker, L.~B., 1978. Evolutionary stable strategies and game
  dynamics. Mathematical biosciences 40~(1), 145--156.

\bibitem{terras}
Terras, A., 1999. Fourier analysis on finite groups and applications. Vol.~43.
  Cambridge University Press.

\bibitem{tilman}
Tilman, D., Kareiva, P.~M., 1997. Spatial ecology: the role of space in
  population dynamics and interspecific interactions. Vol.~30. Princeton
  University Press.

\bibitem{traulsen}
Traulsen, A., Claussen, J.~C., Hauert, C., 2005. Coevolutionary dynamics: from
  finite to infinite populations. Physical review letters 95~(23), 238701.

\bibitem{utz}
Utz, M., Kisdi, {\'E}., Doebeli, M., 2007. Quasi-local competition in
  stage-structured metapopulations: a new mechanism of pattern formation. Bull.
  Math. Biol. 69~(5), 1649--1672.

\bibitem{vickers}
Vickers, G., 1989. Spatial patterns and ess's. Journal of theoretical biology
  140~(1), 129--135.

\bibitem{hauertsp}
Wakano, J.~Y., Hauert, C., 2011. Pattern formation and chaos in spatial
  ecological public goodsgames. Journal of theoretical biology 268~(1), 30--38.

\end{thebibliography}

\end{document}